\documentclass{aa}
\usepackage[varg]{txfonts}
\usepackage{graphicx}%
\DeclareGraphicsExtensions{.pdf,.png,.jpg}
\usepackage{amssymb}
\usepackage{hhline}
\usepackage{tablefootnote}

\begin{document}

    \title{Distance and mass of the M104 (Sombrero) group}
\titlerunning{M 104 (Sombrero)}

\author{Igor D. Karachentsev\inst{1}, 
Lidia N. Makarova\inst{1},
 R. Brent Tully\inst{2}, 
 Gagandeep S. Anand\inst{2},
 Luca Rizzi\inst{3}, and Edward J. Shaya\inst{4} }  
\authorrunning{I.\,Karachentsev, L.\,Makarova, B.\,Tully et al.}
 \institute{Special Astrophysical Observatory, The Russian Academy of Sciences, Nizhnij Arkhyz, Karachai-Cherkessian Republic
369167, Russia
 \and Institute for Astronomy, University of Hawaii,  2680 Woodlawn Drive, Honolulu, HI 96822, USA
\and W. M. Keck Observatory, 65-1120 Mamalahoa Hwy,  Kamuela, HI 96743, USA
 \and Astronomy Department, University of Maryland, College Park, MD 20743, USA}

\date{\today}

\abstract
 {}
{Distances and radial velocities of galaxies in the vicinity
of the luminous  early-type galaxy M\,104 (Sombrero) are used to derive its dark matter
mass.}
{Two dwarf galaxies: UGCA\,307 and KKSG\,30 situated near
M 104 were observed with the Advanced Camera for Surveys on the Hubble
Space Telescope. The distances $9.03^{+0.84}_{-0.51}$ Mpc (UGCA\,307) and 
$9.72^{+0.44}_{-0.41}$ Mpc (KKSG\,30) were determined using the tip of the red giant branch method.
These distances are consistent with the dwarf galaxies being satellites of
  Sombrero.}
{Using radial velocities and projected separations of UGCA\,307, 
KKSG\,30, and a third galaxy with an accurate distance (KKSG\,29), as well as 12 other assumed companions with less accurate distances, the total mass of M\,104 is estimated to be $(1.55\pm0.49) 10^{13} M_{\odot}$. At the $K$-band luminosity of the Sombrero galaxy of $2.4 10^{11} L_{\odot}$, its total mass-to-luminosity ratio is $M_T/L_K = (65\pm20) M_{\odot}/L_{\odot}$, which is about three times higher than that of luminous bulgeless galaxies.}                    
{}

\keywords{galaxies: dwarf --- galaxies: distances and redshifts --- galaxies: photometry --- galaxies: individual: UGCA\,307, KKSG\,30, M\,104}

\maketitle
 \section{Introduction}

  The Local Volume of the Universe amounts to almost a thousand galaxies having distance estimates within 11~Mpc\footnote{http://www.sao.ru/lv/lvgdb} (Karachentsev et al. 2013).
Near the far edge of this volume at a distance of 9.55~Mpc (McQuinn et al. 2016) there is a bright galaxy M\,104 (also known as  NGC\,4594 or the Sombrero galaxy). With an apparent 
  $K$-band magnitude of $m_K = 5\fm0$ it has the
luminosity of $L_K/L_{\odot} = 11.32$~dex, which  is four times higher than the luminosity of the Milky Way (10.70~dex) or M\,31 (10.73~dex). Thanks to  its luminosity and, by inference, 
to its stellar mass the Sombrero is the most
outstanding galaxy of the Local Volume.

  Over recent years several attempts have been undertaken to determine
the total (virial) mass of Sombrero using radial velocities and projected
separations of its companions. Estimations of $M_T/M_{\odot}$ vary widely: 10.90~dex (Makarov \& Karachentsev, 2011), 13.17~dex (Karachentsev \& Nasonova, 2013), 13.45~dex (Karachentsev \& Kudrya, 2014), and 13.96~dex
(Kourkchi \& Tully, 2017). The main reason of the scatter in estimates of
the total mass is caused by the uncertainty on the gravitational binding of Sombrero with galaxies neighbouring in the projection. The Sombrero galaxy is located near the
equator of the Local Supercluster where galaxies are concentrated in the filamental structure, the Virgo Southern Extension (VirgoSE; Tully, 1982;
Kourkchi \& Tully, 2017). Many galaxies in the VirgoSE have radial velocities similar to that of Sombrero, but lie at greater distances typical of the Virgo cluster (15--20~Mpc). 
At a distance of 8 Mpc from the core of the Virgo cluster, the Sombrero galaxy lies at the edge of the zero velocity surface bounding the cluster infall domain, a property shared by other galaxies in the VirgoSE over an extended range in distances.

To reveal the true satellites of Sombrero among the neighbouring galaxies we need to measure their distances, preferably with an error $\Delta D<$\,\,$\sim 1$~Mpc. At present there is only one
galaxy, KKSG\,29, in the Sombrero vicinity with an accurately measured distance (9.82$\pm$0.32~Mpc; Karachentsev et al. 2018), determined via the tip of the red giant branch (TRGB) method. This distance places the dwarf galaxy KKSG\,29 as a physical satellite of the Sombrero galaxy.
 
 In this work we present measurements of TRGB distances for two more  dwarf galaxies,  UGCA\,307 (or DDO\,153) and KKSG\,30 (or LEDA\,3097708) situated close to Sombrero.
The distances of both the galaxies agree well with their belonging to the
family of Sombrero satellites. The new distance measurements together with
other less reliable  distance estimates give us a possibility to make more precise value of the virial (orbital) mass of Sombrero.

 \section{TRGB distances to UGCA\,307 and KKSG\,30}

The dwarf galaxies UGCA\,307 ($12^h53^m56\fs8$--$12\degr06\arcmin21\arcsec$) and KKSG\,30 ($12^h37^m35\fs9$--$08\degr52\arcmin01\arcsec$) have apparent
 $B$ magnitudes of $14\fm6$ and $16\fm3$, respectively, and projected separations of $\sim3\degr$ with respect to Sombrero.
Their radial velocities in the Local Group rest frame,  731~km\,s$^{-1}$ (UGCA\,307)
and 918~km\,s$^{-1}$ (KKSG\,30), are close to the radial velocity of Sombrero, 892~km\,s$^{-1}$. 
  The galaxies were observed with the Advanced Camera for Surveys (ACS) aboard the Hubble Space Telescope (HST) on December 5, 2019, and
March 13, 2020, as a part of the SNAP project 15922 (PI R.B. Tully).

\begin{figure*} 
\begin{tabular}{cc}
\includegraphics[width=\textwidth]{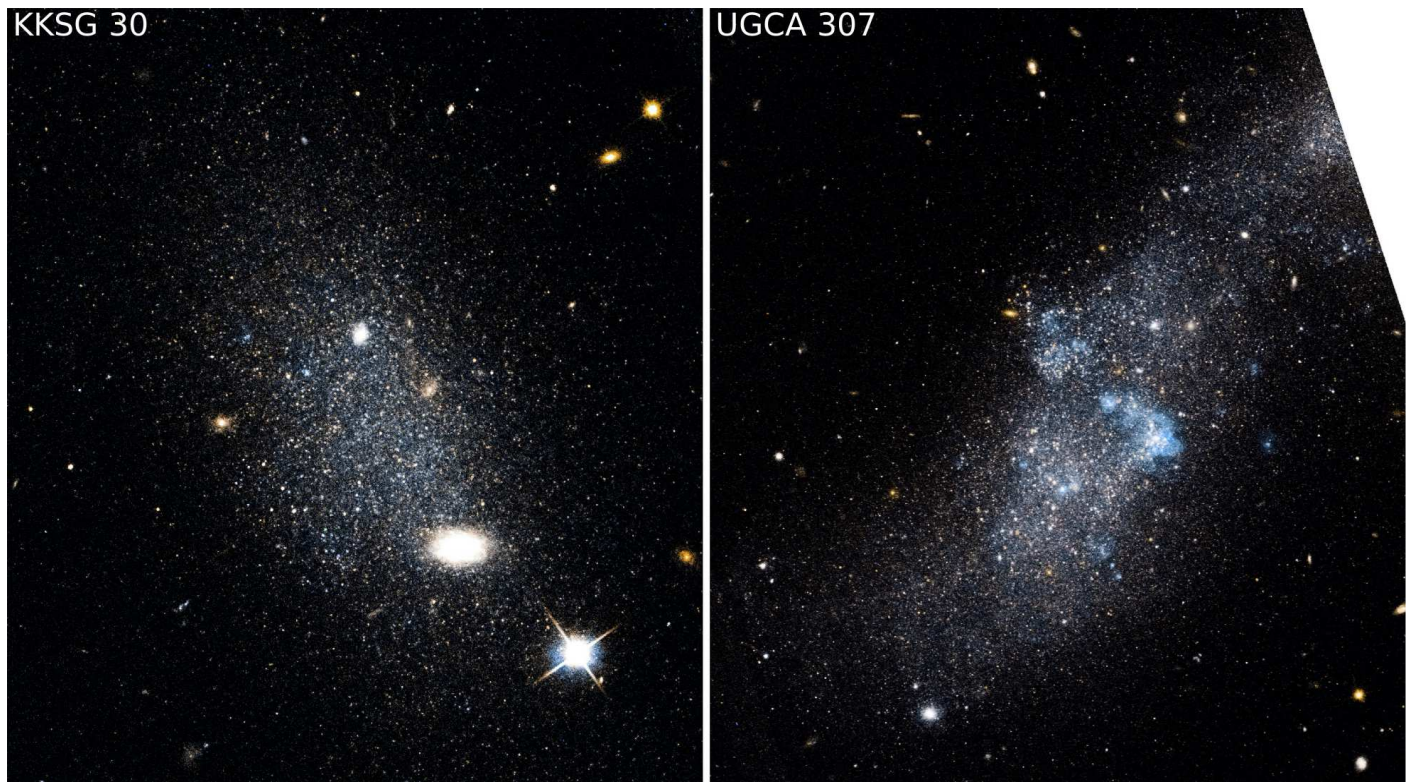}
\end{tabular}
\caption{ HST/ACS combined images of UGCA 307 and KKSG 30.
          The image size is $1\farcm6\times1\farcm4$. North is up and east is left.}
\label{m104:fig01}
\end{figure*}

\begin{figure*} 
\begin{tabular}{cc}
\includegraphics[width=0.5\textwidth]{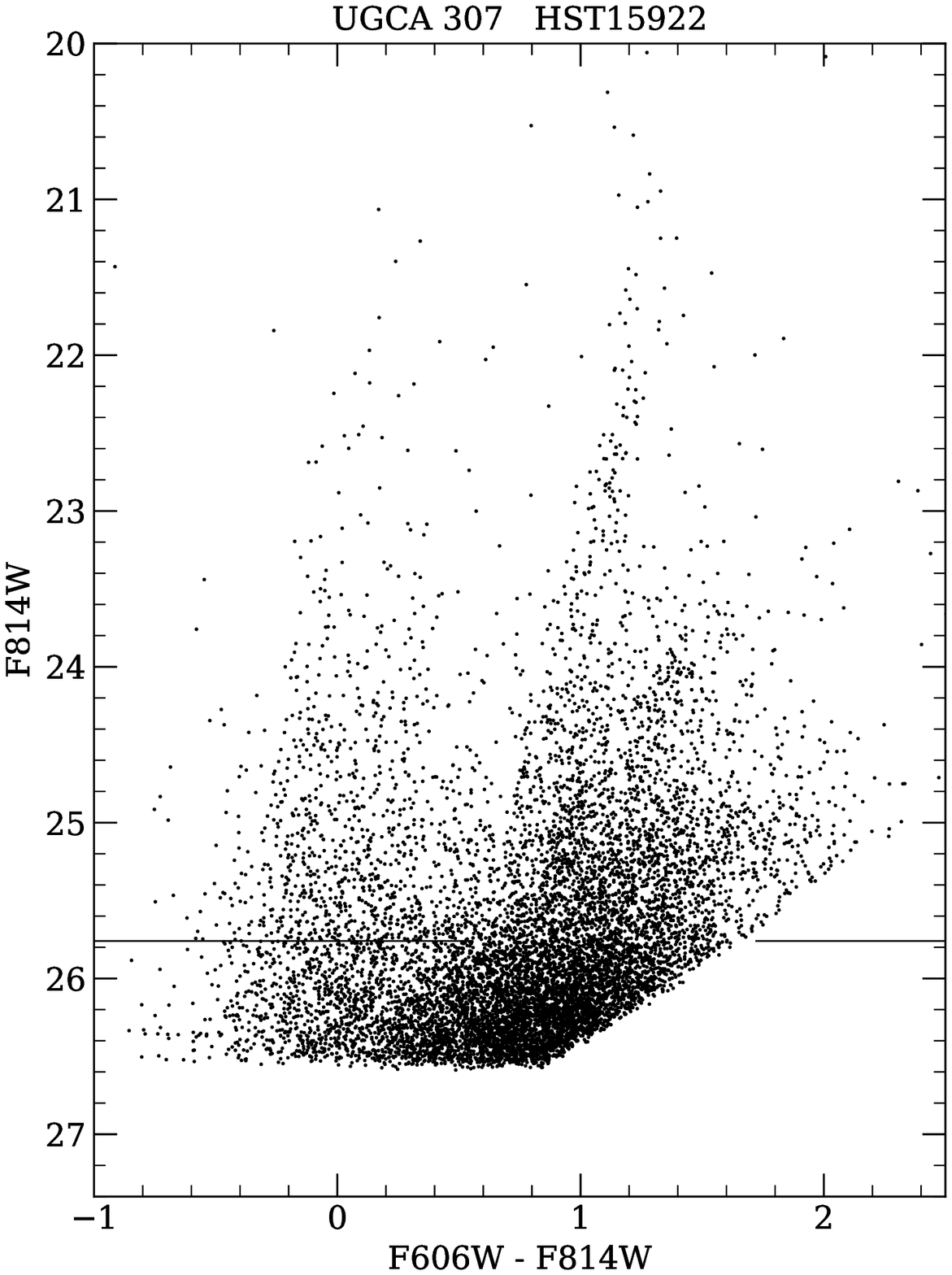}
\includegraphics[width=0.5\textwidth]{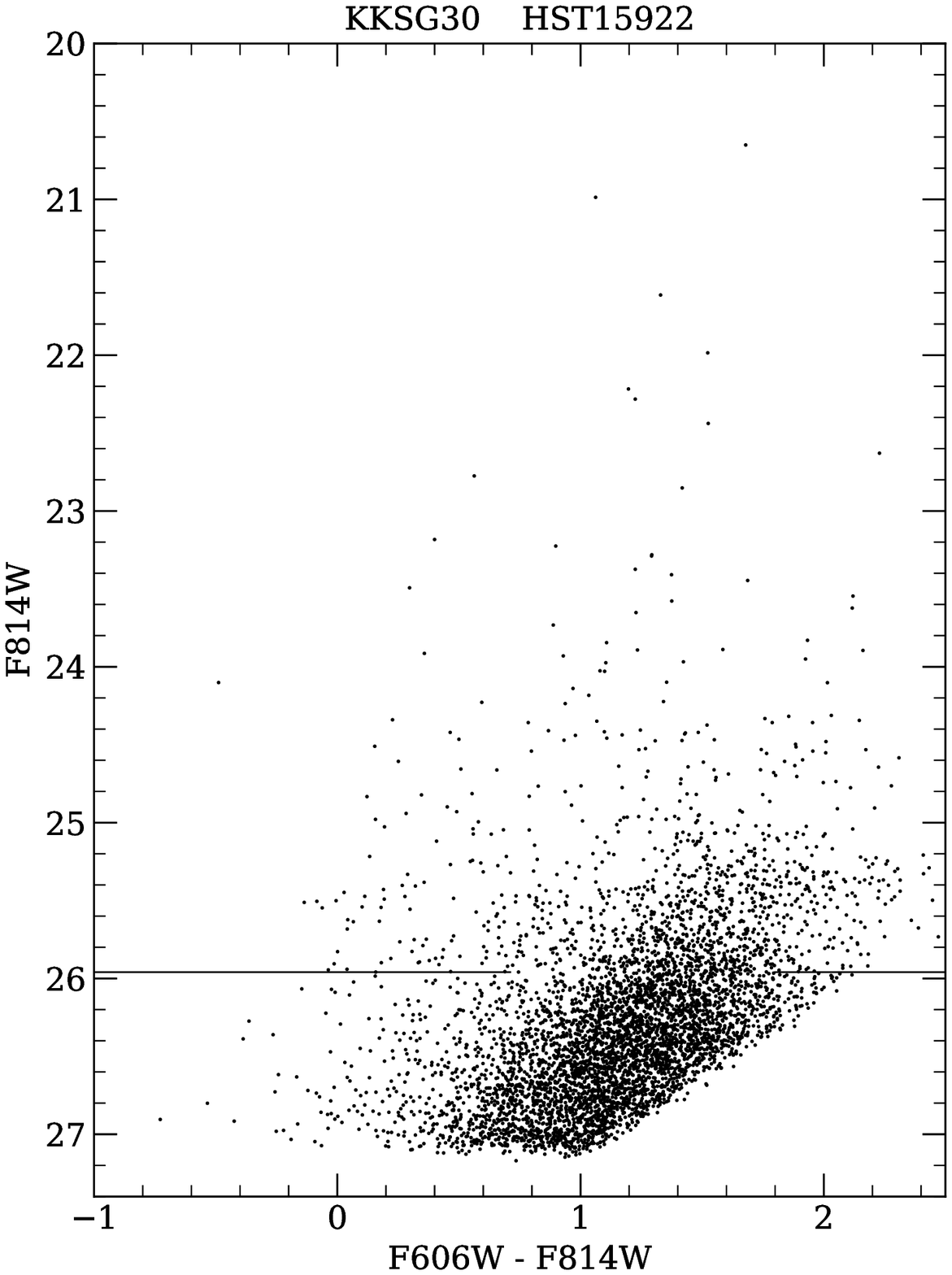}
\end{tabular}
\caption{Colour--magnitude diagrams of UGCA\,307 and KKSG\,30. The TRGB position is indicated by the horizontal line.}
\label{m104:fig02}
\end{figure*}

  Two exposures for each object were made in a single orbit with the filters $F606W$ (750~s) and $F814W$ (750~s). The $F814W$ images of the galaxies are presented
  in Fig.~\ref{m104:fig01}. We used the ACS module of the DOLPHOT package by
Dolphin (2002) to perform photometry of resolved stars based on
the recommended recipe and parameters. Only stars with good quality
photometry were included in the analysis. We selected the stars with a signal-to-noise ratio $S/N > 4$ in  both filters, and with DOLPHOT parameters $crowd_{F606W} + crowd_{F814W} < 0.8$, ($sharp_{F606W}+sharp_{F814W})^2 < 0.075.$ Artificial stars were inserted and recovered using the same reduction procedures to accurately estimate photometric errors. Subsequent analysis included only those image regions that contain stars of the galaxies themselves. We selected the region
of $1.6 \times 1.6$ arcmin around UGCA\,307 and $2.8 \times 1.5$ arcsec around KKSG\,30. The resulting
colour-magnitude diagrams in $F606W$--$F814W$ versus $F814W$ are plotted in Fig.~\ref{m104:fig02}. A maximum-likelihood method by Makarov et al. (2006) was
applied to estimate the magnitude of the TRGB. We found $F814W$(TRGB) to be $25.76^{+0.19}_{-0.11}$ for 
UGCA\,307 and $25.96^{+0.08}_{-0.07}$ for KKSG\,30. Following the zero-point calibration of the absolute magnitude of the TRGB developed by Rizzi et al. (2007), we obtained M(TRGB) values of  --4.09 (UGCA\,307) and --4.08 (KKSG\,30). Assuming values of  foreground reddening, $E(B-V),$ 0.049 (UGCA\,307) and 0.028 (KKSG\,30) from Schlafly \& Finkbeiner (2011), we
derived the true distance modulus of $(m- M)_0 = 29.78^{+0.20}_{-0.12}$ or
the distance $D = 9.03^{+0.84}_{-0.51}$~ Mpc for UGCA\,307 and $(m- M)_0 = 29.94^{+0.10}_{-0.09}$
or the distance $D = 9.72^{+0.44}_{-0.41}$ Mpc for KKSG\,30.

 \section{Companions of Sombrero and background objects}
\begin{table*}
\caption{Galaxies around Sombrero with $r_p < 6\degr$ and $V_{LG} <1400$ km s$^{-1}$.}
\label{m104:tab01}
\begin{tabular}{lcrlrrlc} \hline

 Name       &       RA  Dec  &  $V_{LG}$& Type&  $B_T$  &  $D$   & meth. &Ref.    \\
\hline   
(1)& (2)& (3)& (4)& (5)& (6)& (7)& (8) \\   \hline                                                     
DDO116      & 121628.6$-$113141&   959& Sm  & 16.02 & 23.28&TFb$^1$& (1)  \\
DDO118      & 121711.9$-$114041&  1064& Irr & 16.2  & 14.22&  TFb  & (1)  \\
PGC104765   & 122143.7$-$123957&  1221& Irr & 16.92 &      &       &      \\
KKSG27      & 122205.7$-$094801&  1128& Im  & 17.70 &  6.64&  TFb  & (1)  \\
UGCA278     & 122310.4$-$135644&   942& Irr & 16.20 & 18.2 &  TF   & (1)  \\
PGC970571   & 122549.4$-$110305&  1134& BCD & 16.33 &      &       &      \\
PGC157820   & 123001.8$-$114731&   909& S0  & 16.13 &      &       &      \\
NGC4487     & 123104.5$-$080314&   847& Sc  & 11.76 & 20.1 &  SN   & (2)  \\
NGC4504     & 123217.4$-$073348&   812& Sc  & 12.12 & 17.5 &  TF   & (3)  \\
UGCA287     & 123355.4$-$104047&   852& Sm  & 15.36 & 20.5 &  TF   & (4)  \\
UGCA289     & 123537.8$-$075236&   800& Sd  & 14.46 & 14.6 &  TF   & (3)  \\
PGC970397   & 123539.4$-$110402&   928& Irr & 16.87 & 10.00&  TF   & (1)  \\
KKSG29      & 123714.1$-$102952&   562& Irr & 16.54 &  9.82&  TRGB & (5)  \\
KKSG30      & 123735.9$-$085201&   918& Irr & 16.30 &  9.72&  TRGB & (8)  \\
dw1239-1143 & 123915.3$-$114308&  1171& dE  & 16.80 &  7.9 &  SBF  & (6)  \\
PGC1024539  & 123944.7$-$070519&   744& Irr & 17.90 & 15.6 &  TFb  & (1)  \\
NGC4594     & 123959.4$-$113723&   892& S0a &  9.00 &  9.55&  TRGB & (7)  \\
SUCD1       & 124003.1$-$114004&  1109& dE  & 18.40 &  9.55&  mem  & (1)  \\
NGC4597     & 124012.9$-$054757&   866& Sm  & 12.87 & 10.10&  TF   & (4)  \\
dw1240-1140 & 124017.6$-$114046&  1097& dSph& 19.50 &  9.55&  mem  &      \\
PGC042730   & 124248.9$-$122327&   829& dEn & 14.78 &  9.55&  mem  &      \\
UA295       & 124453.9$-$090731&  1197& Sm  & 15.10 & 22.9 &  TF   & (1)  \\
DDO146      & 124541.4$-$060408&  1304& Sm  & 13.01 & 17.3 &  TF   & (3)  \\
NGC4674     & 124603.5$-$083920&  1318& Sab & 13.96 &      &       &      \\
PGC1003283  & 124750.9$-$082816&   860& S0  & 15.89 &      &       &      \\
PGC104868   & 124854.1$-$114042&  1171& BCD & 15.0  & 11.17&  TF   & (1)  \\
NGC4700     & 124908.1$-$112435&  1219& Sd  & 12.60 &  7.30&  TF   & (4)  \\
PGC043345   & 124923.6$-$100712&  1124& Sdm & 12.79 & 16.6 &  TF   & (4)  \\
PGC1019240  & 124955.9$-$072527&  1236& Sm  & 15.52 &      &       &       \\
NGC4731     & 125101.4$-$062339&  1323& SBc & 11.88 & 13.6 &  TF   & (3)    \\
NGC4723     & 125103.0$-$131413&  1109& Sm  & 15.38 & 15.3 &  TF   & (4)    \\
PGC043526   & 125113.3$-$063334&  1327& Im  & 13.26 & 13.6 &  mem  &        \\
NGC4742     & 125148.0$-$102717&  1088& E   & 12.11 & 15.8 &  SBF  & (9)    \\
NGC4757     & 125250.1$-$101836&   664& S0  & 14.54 &      &       &        \\
UGCA 307    & 125356.8$-$120621&   731& Im  & 14.60 &  9.03&  TRGB & (8)    \\
NGC4781     & 125423.7$-$103214&  1080& Scd & 11.39 & 15.5 &  TF   & (4)  \\
NGC4790     & 125451.9$-$101452&  1175& Sc  & 12.41 & 16.9 &  TF   & (4)  \\
UGCA308     & 125531.1$-$102350&  1140& Irr & 16.30 & 16.6 &  TFb  & (4)  \\
NGC4802     & 125549.6$-$120319&   843& S0  & 12.39 & 11.5 &  SBF  & (10) \\
IC3908      & 125640.6$-$073346&  1127& Scd & 13.33 & 23.9 &  TF   & (4)  \\
NGC4818     & 125648.9$-$083131&   892& Sab & 11.99 & 11.3 &  TF   & (3)  \\
UGCA311     & 125746.8$-$093802&  1306& Scd & 14.00 & 19.8 &  TF   & (4)  \\
PGC044460   & 125828.3$-$103437&  1173& Sdm & 14.70 &  8.70&  TF   & (4)\\
UGCA312     & 125906.5$-$121340&  1121& Irr & 15.88 & 12.0 &  TFb  & (4)\\
NGC4856     & 125921.3$-$150232&  1145& S0a & 11.44 & 24.0 &  TF   & (4)\\
UGCA314     & 130017.2$-$122048&  1397& Im  & 14.64 & 24.8 &  TF   & (4)\\
PGC936912   & 130107.0$-$133106&  1120& Im  & 15.40 & 14.6 &  TF   & (4)\\
NGC4920     & 130204.2$-$112243&  1155& Im  & 14.15 & 18.2 &  TF   & (4)\\
\hline 
                                                                
\multicolumn{8}{p{0.5\textwidth}}{(1)  Kashibadze+2018, (2)  Pejcha+2015, (3)  Tully+2016, (4)  Karachentsev+2013, (5)  Karachentsev+2018, (6)  Carlsten+2020, (7)  McQuinn+2016, (8)  present paper, (9)  Blakeslee+2001, (10)  Tonry+2001.} \\
\multicolumn{8}{p{0.5\textwidth}}{$^1$ "TRGB" -- the luminosity of tip of the red giant branch, "SBF" -- surface brightness fluctuations,"TF" and "TFb" -- the classic Tully \& Fisher (1977) relation or its baryonic version; "SN" -- the luminosity of supernova; "mem" -- assumed membership in a group with the known distance}
\end{tabular}
 \end{table*}

  Judging by the big stellar mass of the Sombrero galaxy, the virial radius of its halo can reach about 400~kpc. To search for assumed satellites of Sombrero
we examined a region of radius $r_p = 6\degr$ around it that corresponds to
the linear projected radius of $R_p = 1.0$~Mpc at the distance of 9.55~Mpc.
In this area there are 48 galaxies with radial velocities $V_{LG} < 1400$~km\,s$^{-1}$.
The data   are presented in Table~\ref{m104:tab01}. The table columns contain  (1) galaxy
name; (2) equatorial coordinates J2000.0; (3) radial velocity in the Local
Group rest frame (km\,s$^{-1}$); (4) morphological type; (5) apparent $B$ magnitude from the Lyon Extragalactic Database (LEDA, Makarov et al. 2014) or NASA Extra-galactic Database (NED); (6) distance to the galaxy in Mpc; (7) method used
for the distance estimate; (8) reference to the source of distance.

\begin{figure} 
\centering
\includegraphics[width=\hsize]{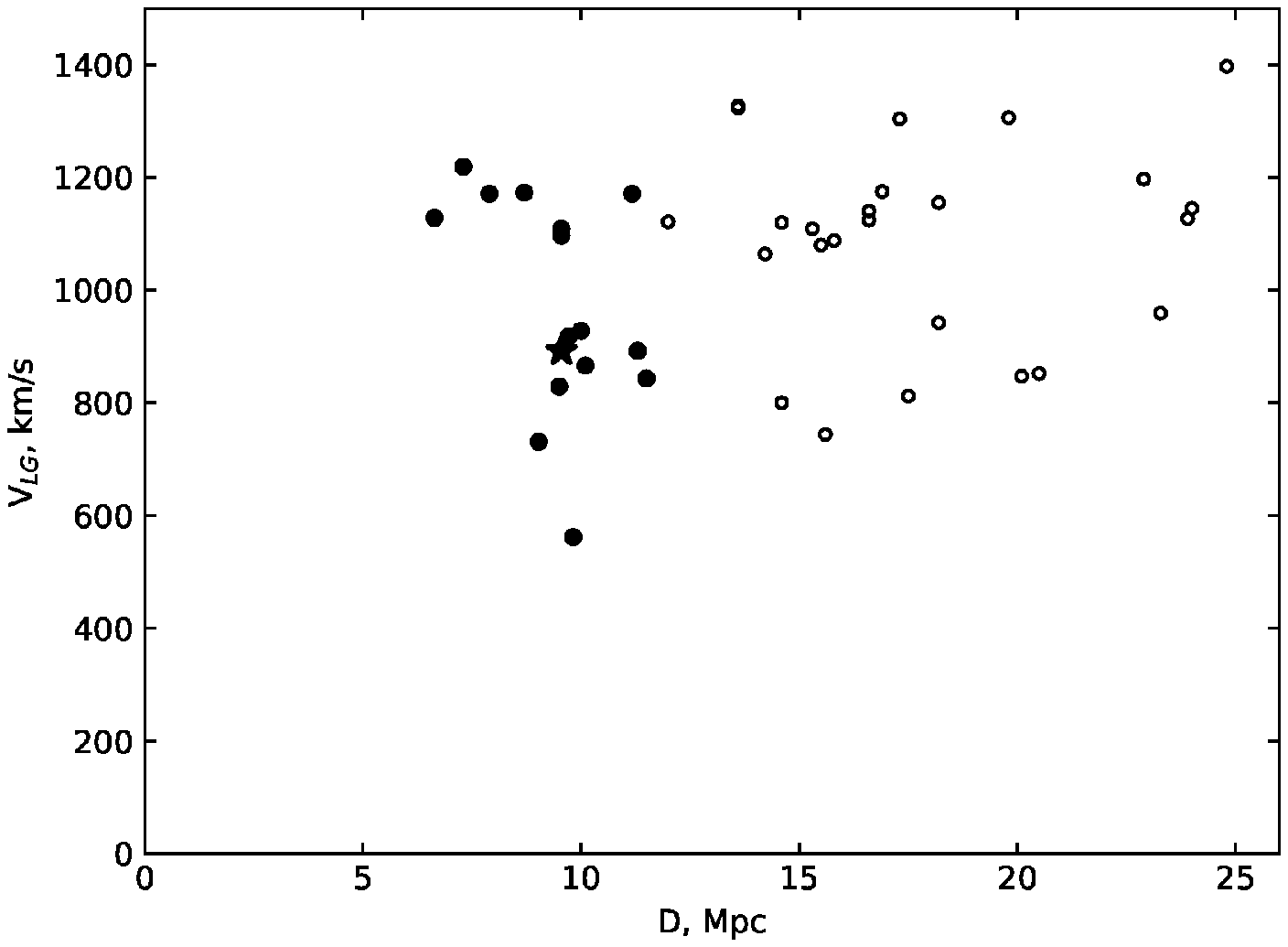}
\caption{ Distribution of assumed members of the Sombrero group (filled circles) and background galaxies (open circles) according to their 
distance and radial velocity in the Local Group rest frame. The galaxies satisfy the conditions $V_{LG} < 1400$~km\,s$^{-1}$ and projected 
separation $r_p < 6\degr$ with respect to  Sombrero (asterisk).}
\label{m104:fig03}
\end{figure}

  As seen from these data, 41 of the 48 galaxies  have distance estimates. Most of them were made by the Tully-Fisher method, with uncertainties of 35-30\% for these 
low luminosity galaxies. Accordingly, we consider
only galaxies with distances $D < 12$~Mpc as probable members of the Sombrero
group. The distance and radial velocity distribution of galaxies around Sombrero are given in Figure~\ref{m104:fig03}. In total, 15 galaxies are
probable satellites of Sombrero, with the luminosity of each of them
more than an order of magnitude fainter than the luminosity of Sombrero.
An empty volume (mini-void) is visible ahead of the group. The background
galaxies have radial velocities substantially overlapping  the Sombrero
group velocities. Due to significant TF distance errors, $\Delta$D of about 3-5~Mpc,
the membership of some galaxies (UGCA\,312, PGC\,104868), whether in the group or background may be subject to revision.

\begin{figure} 
\centering
\includegraphics[width=\hsize]{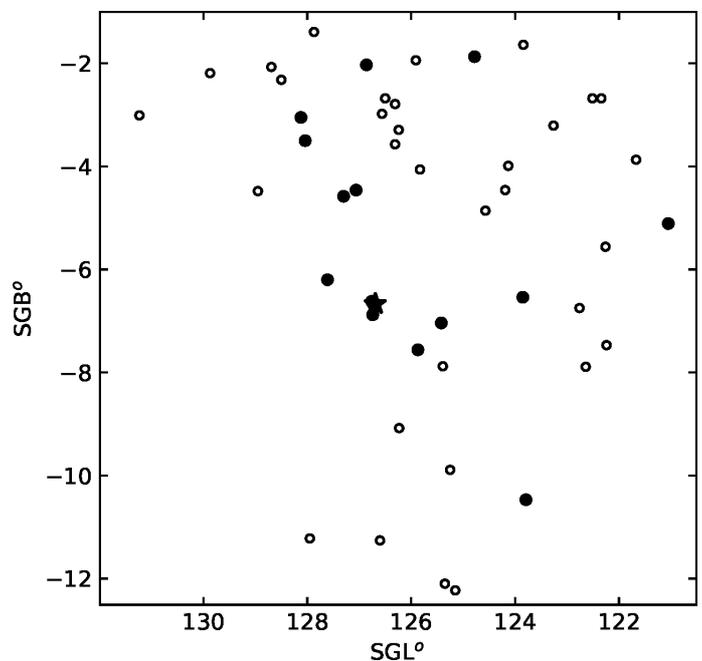}
\caption{Distribution of assumed Sombrero satellites (filled circles) and background galaxies (open circles) in supergalactic coordinates. 
 All galaxies have $V_{LG} < 1400$~km\,s$^{-1}$.}
\label{m104:fig04}
\end{figure}
  The distribution of galaxies from Table~\ref{m104:tab01} in supergalactic coordinates SGL, SGB
is presented in Figure~\ref{m104:fig04}. Indications of galaxies with different symbols are the same as in the previous figure. Sombrero's satellites, as well as
 background galaxies, are distributed asymmetrically. In both subsamples there is
a noticeable increase in galaxy number towards the supergalactic equator and
towards the Virgo cluster centre (SGL = 103$\degr$, SGB = $-2 \degr$). The reason for this asymmetry in the case of Sombrero group members remains unclear to us.
 
\begin{figure} 
\centering
\includegraphics[width=\hsize]{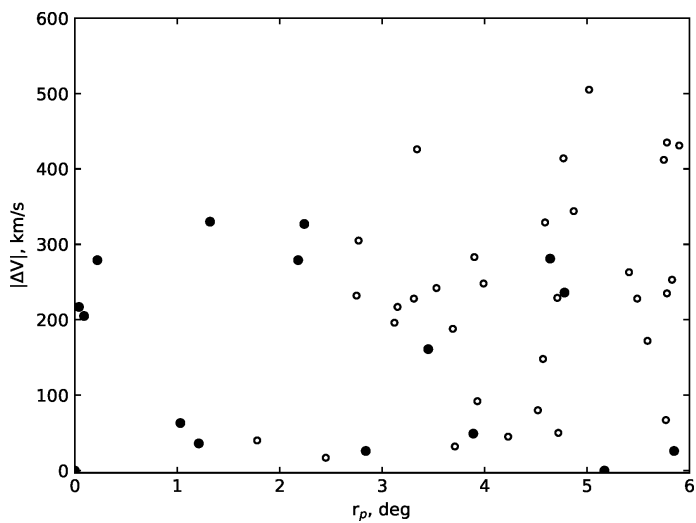}
\caption{ Radial velocity difference and angular projected separation for assumed members of the Sombrero group (filled circles) and background
 galaxies (open circles) taken with respect to the Sombrero galaxy.}
\label{m104:fig05}
\end{figure}

 Figure~\ref{m104:fig05} presents the distribution of assumed satellites of Sombrero (solid
circles) and background galaxies (open circles) according to angular projected separation, $r_p,$ and absolute value of radial velocity difference,
$|\Delta V|$. Sombrero satellites dominate within $r_p < 2.4\degr$ (i.e. 400~kpc),
and at larger distances assumed Sombrero satellites are lost among the numerous
background galaxies. Such a confusion of two categories of galaxies makes it
difficult to estimate the virial mass of the Sombrero galaxy.

\section{Total (orbital) mass of  Sombrero}
\begin{table}
 \caption{Satellites of Sombrero with known radial velocities.}
\label{m104:tab02}
\begin{tabular}{llcrr}\hline
  Name      &   Type&    $r_p$,&   $R_p$,&   $\Delta V$ \\
\hline                                       
                    &        & $\degr$ &   kpc  &km s$^{-1}$   \\
\hline                                       
 SucD1      &   dE  &  0.04  &   7  & 217   \\
 dw1240-1140&   dSph&  0.09  &  15  & 205   \\
 dw1239-1143&   dE  &  0.22  &  37  & 279   \\
 PGC042730  &   dE  &  1.03  & 171  & -63   \\
 PGC970397  &   Irr &  1.21  & 201  &  36   \\
 KKSG29     &   Irr &  1.32  & 220  &-330   \\
 PGC104868  &   BCD &  2.18  & 362  & 279   \\
 NGC4700    &   Sd  &  2.24  & 372  & 327   \\
 KKSG30     &   Irr &  2.84  & 471  &  26   \\
 UGCA 307   &   Im  &  3.45  & 573  &-161   \\
 NGC4802    &   S0  &  3.89  & 646  & -49   \\
 PGC044460  &  Sdm  & 4.64   &770   &281    \\
 KKSG27     &   Im  &  4.78  & 794  & 236   \\
 NGC4818    &   Sab &  5.17  & 858  &   0   \\
 NGC4597    &   Sm  &  5.85  & 971  & -26   \\
\hline\end{tabular}
\end{table}

  The list of 15 assumed satellites of Sombrero with known radial velocities is presented in Table~\ref{m104:tab02}. The galaxies are ranked according to their
  projected separation from Sombrero. The average linear projected separations of the satellites is $\langle R_p\rangle $ = 431$~kpc $, the mean difference of
their radial velocities with respect to the principal galaxy is $+62\pm54$~km\,s$^{-1}$,
and the dispersion of radial velocities is $\sigma_v = 204$~km\,s$^{-1}$.
 
 We estimated the virial (orbital) mass of the Sombrero galaxy assuming Keplerian
motion of satellites, as test particles, around the massive central body.
At random orientation of the satellite orbits with the mean eccentricity
of $\langle e^2\rangle = 1/2$ (Barber et al. 2014) the estimate of orbital mass can be
written (Karachentsev \& Kudrya, 2014) as

\begin{equation}
    M_{\rm orb} = (16/\pi) G^{-1} \langle \Delta V^2 R_p\rangle,
\end{equation}

where $G$ is the gravitational constant. If $R_p$ is expressed in kpc and
$\Delta V$ expressed in km\,s$^{-1}$, then

\begin{equation}
    \log(M_{\rm orb}/M_{\odot}) = \log \langle\Delta V^2 R_p\rangle + 6.07.
\end{equation}

Using all 15 assumed satellites from Table 2 we obtain the mean value of orbital 
mass to be  $(17.2\pm6.1) 10^{12} M_{\odot}$. For three satellites with accurate TRGB distances
this quantity is $(15.3\pm8.1) 10^{12} M_{\odot}$, while for five dwarfs with TRGB and SBF distances
the average orbital mass drops to $(10.3\pm5.4) 10^{12} M_{\odot}$. Attributing to 15 satellite
galaxies different weights (w=1 for TRGB distances, w= 1/2 for SBF distances and w= 1/4 for TF
and mem distances), we derive the weighted mean $(15.5\pm4.9) 10^{12} M_{\odot}$.
We adopt this  quantity   as the optimal estimate of the total mass of the Sombrero group.

  As seen from Table~\ref{m104:tab02}, the early-type galaxies are concentrated towards
Sombrero much more tightly than spiral and irregular galaxies. Apart from the S0 galaxy
NGC\,4802, all other E-companions reside in the central zone $R_p < 200$~kpc.
The known effect of morphological segregation is more pronounced if
probable Sombrero satellites without radial velocities are taken into
account. Table~\ref{m104:tab03} lists 12 dwarf galaxies of low and very low surface brightness have been detected near Sombrero
 by different authors (Karachentsev et al. 2000; Javanmardi et al. 2016; Karachentsev et al. 2020;
Carlsten et al. 2020). None of them is detected in the HI line, and all are
classified as spheroidal dwarfs. These objects with quenched star formation
have projected separations $R_p < 200$~kpc, increasing the segregation effect.
\begin{table}
\caption{Assumed satellites of Sombrero 
           without radial velocities.}
\label{m104:tab03}
\begin{tabular}{lclrl} \hline
  Name         & RA (2000.0) DEC & Type &D, Mpc & meth\\
\hline                                  
 dw1237-1125   & 123711.6-112559 & dSph & 7.5 & SBF \\
 KKSG31        & 123833.7-102925 & dSph & 9.55& mem \\
 dw1239-1152   & 123909.0-115237 & dSph & 8.2 & SBF \\
 dw1239-1159   & 123909.1-115912 & dSph &11.3 & SBF \\
 N4594-DGSAT-3 & 123932.8-111338 & dSph & 7.9 & SBF \\
 Sombrero DwA  & 123951.5-112029 & dSph & 9.7 & SBF \\
 KKSG32        & 123955.0-114448 & dSph & 9.0 & SBF \\
 KKSG33        & 124008.9-122153 & dSph & 9.55& mem \\
 dw1240-1118   & 124009.4-111850 & dSph & 8.8 & SBF \\
 dw1241-1131   & 124102.8-113144 & dSph & 7.2 & SBF \\
 Sombrero DwB  & 124112.0-115333 & dSph &11.2 & SBF \\
 KKSG34        & 124118.9-115539 & dSph & 9.0 & SBF \\
\hline\end{tabular}
\end{table}
  \section{Concluding remarks}

  Radial velocities and projected separations of 15 assumed satellites of Sombrero yield the weighted estimate of its total mass 
$(M_T/M_{\odot})= (1.55\pm0.49)10^{13}$.
At $M*/L_K = 1 M_{\odot}/L_{\odot}$ (Bell et al. 2003) the stellar mass of Sombrero is $2.1\times10^{11} M_{\odot}$. Accounting for  the luminosity of all the
 satellites increases the stellar mass of the group to $M* = 2.4\times 10^{11} M_{\odot}$. Therefore, the Sombrero halo has a total-mass-to-stellar-mass ratio of
$M_T/M* = 65\pm20$, which is much higher than the cosmic baryonic ratio, $M_{\rm halo}/M_b = 6$.
 
 Karachentseva et al. (2011) undertook a search for faint companions around
2MASS Isolated Galaxies (2MIG). They found 214 faint neighbours around 125 2MIG galaxies with radial velocity differences of $|\Delta V| < 500$~km\,s$^{-1}$ and projected
 separations of $R_p < 500$~kpc. For 60 companions around E,S0-galaxies
the median ratio of $M_{\rm orb}/M*$ turns out to be 63, while for the remaining 154 spiral galaxies the median ratio is only 17. A similar search for companions
around late-type spiral galaxies without bulges was performed by Karachentsev \& Karachentseva (2019). Based on 43 companions around 30 Sc-Scd-Sd galaxies, they
 found the mean ratio $M_{\rm orb}/M* = 20\pm3.$ The factor of three difference in
halo-mass-to-stellar-mass ratio between early-type and late-type luminous galaxies attests to their different dynamical histories.
  
In the vicinity of Sombrero there are still more than a dozen galaxies
with unreliable or even unknown distance estimates. Measurements of their TRGB distances with HST would help us to study the structure and dynamics of this group.

\begin{acknowledgements}
We are grateful to the referee for helpful advice.
   This work is based on observations made with the NASA/ESA
Hubble Space Telescope. STScI is operated by the Association of Universities for Research in Astronomy, Inc. under NASA contract NAS 5-26555. The work in Russia is 
supported by RFBR grant 18-02-00005.
\end{acknowledgements}

{}
\end{document}